# A quantitative first-order approach for the scattering of light by structured thin films

**SLAWA LANG,[1] LUKAS MAIWALD,[1,*] HAGEN RENNER,[1] DIRK JALAS,[1] ALEXANDER YU. PETROV[1,2] AND MANFRED EICH[1,3]**

[1]*Institute of Optical and Electronic Materials, Hamburg University of Technology, Eissendorfer Strasse 38, 21073 Hamburg, Germany*
[2]*ITMO University, 49 Kronverkskii Ave., 197101, St. Petersburg, Russia*
[3]*Institute of Materials Research, Helmholtz-Zentrum Geesthacht, Max-Planck-Strasse 1, Geesthacht, D-21502, Germany*
*\*lukas.maiwald@tuhh.de*

**Abstract:** We present a full vectorial first-order approach to the scattering by arbitrary photonic structures with a low refractive index contrast. Our approach uses the first-order Born approximation and keeps the simple geometrical representation of the Ewald sphere construction. Via normalization to a representative sample volume, the approach can also predict the scattering by infinitely extended layers of scattering media. It can therefore be used to describe and efficiently calculate the scattering by structures where the linear first-order scattering terms dominate, e.g. in low index contrast disordered structures creating a color impression.

## 1. Introduction

The field of disordered photonics has attracted broad interest in recent years [1]. There are many applications based on multiple scattering, e.g. localization [2], random lasing [3] or broadband reflection [4]. Other interesting applications like structural color can adequately be described by first-order scattering [5–7].

While there is a lot of published work on light propagation in ordered photonic structures [8], disordered structures are theoretically less well understood. Furthermore, due to the lack of geometrical order, the simulation of these structures often requires large volumes which cannot be handled easily.

In the following, we present a full vectorial approach to calculate the scattering occurring at an arbitrary spatial perturbation of the permittivity $\varepsilon$, and in particular the first-order term that approximates the full scattered field for weak permittivity perturbations. The approach is similar to the scalar first-order Born approximation which is commonly used in X-ray scattering [9,10] and sometimes applied to optics [11,12]. We show that the Ewald sphere construction directly follows from the Fourier transform of the general solution for the scattered field, which is a convolution of Green's function with the spatial permittivity perturbation. The full vectorial consideration allows taking polarization effects into account.

We also expand the approach to structures that are infinitely extended in one or two spatial dimensions. This opens up its application to film geometries that are found e.g. in structural color coatings [13-15]. For such films the approach for the calculation of a reflectivity spectrum is shown which allows the direct comparison of our approach to real-life measurements.

## 2. First-order scattering by finite perturbation

We assume a nonmagnetic, lossless, source-free material and a monochromatic dependence $\exp(-i\omega t)$ on time $t$, where $\omega$ is the angular frequency of the light. The total electric field $\vec{E}$ as a function of spatial position $\vec{r}$ is then governed by the partial differential equation

$$\nabla \times \nabla \times \vec{E}(\vec{r}) = \frac{\omega^2}{c_0^2} \varepsilon(\vec{r}) \vec{E}(\vec{r}), \tag{1}$$

which is directly derived from Maxwell's curl equations and the material equations [16]. Here, $c_0$ is the vacuum speed of light and $\varepsilon(\vec{r})$ is the relative permittivity. The latter can now be split into the unperturbed, constant permittivity $\bar{\varepsilon}$ and the perturbation $\hat{\varepsilon} f(\vec{r})$, which we further split into the perturbation amplitude $\hat{\varepsilon}$ and the normalized perturbation distribution $f(\vec{r})$. Further, we expand the total electric field in a Taylor series with respect to the perturbation amplitude $\hat{\varepsilon}$:

$$\vec{E} = \vec{E}_0 + \hat{\varepsilon} \vec{E}_1 + \hat{\varepsilon}^2 \vec{E}_2 + \ldots =: \sum_{j=0}^{\infty} \vec{E}_j. \tag{2}$$

Substituting $\varepsilon(\vec{r}) = \bar{\varepsilon} + \hat{\varepsilon} f(\vec{r})$ and (2) into (1) we obtain a set of differential equations

$$\nabla \times \nabla \times \vec{E}_0(\vec{r}) - k^2 \vec{E}_0(\vec{r}) = 0,$$
$$\nabla \times \nabla \times \vec{E}_j(\vec{r}) - k^2 \vec{E}_j(\vec{r}) = \frac{\omega^2}{c_0^2} \hat{\varepsilon} f(\vec{r}) \vec{E}_{j-1}(\vec{r}), \quad j = 1, 2, \ldots, \qquad (3)$$

where we defined $k^2 := \bar{\varepsilon} \omega^2 / c_0^2$. The zeroth-order term $\vec{E}_0$ describes the unperturbed wave which is obviously the incident wave. For simplicity we assume a plane wave $\vec{E}_0(\vec{r}) = \hat{E}_0 \vec{e}_0 \exp(i \vec{k}_0 \vec{r})$ with the unit polarization direction $\vec{e}_0$ lying in the plane perpendicular to $\vec{k}_0$. Other forms of the incident wave can be described as a superposition of plane waves. The higher-order terms $\vec{E}_j$ ($j = 1, 2, \ldots$) describe the scattered field. They can be regarded as waves in the unperturbed medium emitted by sources that are caused by the previous-order field $\vec{E}_{j-1}$ and the permittivity perturbation $\hat{\varepsilon} f(\vec{r})$.

Now, we focus on the first-order term $\vec{E}_1$. It approaches the total scattered field as $\hat{\varepsilon}$ approaches zero. The higher order terms can be treated analogously as seen from (3). While the first-order term alone cannot describe subsequent, multiple scattering events, we note that – depending on the perturbation strength $\hat{\varepsilon}$ – several orders could be required to describe a single scattering event.

To solve the differential equation (3) governing $\vec{E}_1(\vec{r})$, we first find the solution for a point source. This solution is Green's function $\mathbf{G}$ which can be used to find the desired field by convolution:

$$\vec{E}_1(\vec{r}) = \mathbf{G}(\vec{r}) * \frac{\omega^2}{c_0^2} \hat{\varepsilon} f(\vec{r}) \vec{E}_0(\vec{r}) = \hat{E}_0 \frac{\omega^2}{c_0^2} \mathbf{G}(\vec{r}) * \vec{e}_0 \hat{\varepsilon} f(\vec{r}) \exp(i \vec{k}_0 \vec{r}). \qquad (4)$$

We are now exploiting the fact that this convolution in the real space transforms into a simple multiplication in the reciprocal space. As introduced before, the excitation is produced with the plane wave $\vec{E}_0(\vec{r}) = \hat{E}_0 \vec{e}_0 \exp(i \vec{k}_0 \vec{r})$, or $\vec{E}_0(\vec{k}) = (2\pi)^3 \hat{E}_0 \vec{e}_0 \delta(\vec{k} - \vec{k}_0)$ in reciprocal space, which imprints defined phases on the distributed point sources. The product on the right side of (4) $f(\vec{r}) \vec{E}_0(\vec{r}) \propto f(\vec{r}) \exp(i \vec{k}_0 \vec{r})$ turns into a convolution in Fourier space $F(\vec{k}) * \vec{E}_0(\vec{k})/(2\pi)^3 \propto F(\vec{k}) * \delta(\vec{k} - \vec{k}_0)$ resulting in $\hat{E}_0 \vec{e}_0 F(\vec{k} - \vec{k}_0)$, which can be understood as a shift of the Fourier transform of $f(\vec{r})$ in reciprocal space by vector $\vec{k}_0$. The scattered field in real space is given by the inverse Fourier transform:

$$\vec{E}_1(\vec{r}) = \frac{1}{(2\pi)^3} \iiint \vec{E}_1(\vec{k}) \exp(i \vec{k} \vec{r}) d^3 k = \frac{\hat{E}_0}{(2\pi)^3} \frac{\omega^2}{c_0^2} \iiint \mathbf{G}(\vec{k}) \vec{e}_0 \hat{\varepsilon} F(\vec{k} - \vec{k}_0) \exp(i \vec{k} \vec{r}) d^3 k. \qquad (5)$$

It remains to find Green's function in reciprocal space $\mathbf{G}(\vec{k})$ which is done in detail in Appendix A. Having found $\mathbf{G}(\vec{k})$ (see Eq. (A14)) we actually perform the integration in (5) along the $k_z$-direction.

$$\vec{E}_1(\vec{r}) = \frac{\hat{E}_0}{(2\pi)^3} \frac{\omega^2}{c_0^2} \iint \int \mathbf{G}(\vec{k}) \vec{e}_0 \hat{\varepsilon} F(\vec{k} - \vec{k}_0) \exp(i \vec{k} \vec{r}) dk_z d^2 k_\rho$$
$$= \frac{i \hat{E}_0}{2^3 \pi^2} \frac{\omega^2}{c_0^2} \iint \hat{\varepsilon} F(\vec{k}_1 - \vec{k}_0) \left[ (\vec{e}_0 \cdot \vec{e}_1^s) \vec{e}_1^s + (\vec{e}_0 \cdot \vec{e}_1^p) \vec{e}_1^p \right] \frac{\exp(i \vec{k}_1 \vec{r})}{k_{1,z}} d^2 k_{1,\rho}. \qquad (6)$$

The z-component of the scattered wave vector $\vec{k}_1$ is $k_{1,z} = \pm(k^2 - k_{1,\rho}^2)^{1/2}$, where the sign is chosen such that the wave travels from the scattering point towards the observation point $\vec{r}$, and its tangential component is $k_{1,\rho}$. Furthermore, $\vec{e}_1^s := \vec{k}_1 \times \vec{e}_z / |\vec{k}_1 \times \vec{e}_z|$, $\vec{e}_1^p := \vec{k}_1 \times \vec{e}_1^s / |\vec{k}_1 \times \vec{e}_1^s|$ and $\vec{e}_z$ is the unit vector in z-direction. The integration allows us to interpret the solution as a superposition of plane waves, propagating (real $k_{1,z}$) and evanescent ones (imaginary $k_{1,z}$), existing in the unperturbed medium. (Note that in case of evanescent modes the Fourier transform must be extended to complex $\vec{k}$.) The scattered plane waves exist only in a half space ("half space plane waves") because their origin is the scattering perturbation and all scattered waves propagate away from it.

Neglecting evanescent modes, which do not transport energy away from the scatterer, and rewriting the integral in (6), we end up with the first-order scattered field caused by the perturbation. We only need to consider waves propagating towards the observation point $\vec{r}$. Half of the waves are scattered away from $\vec{r}$ and are thus not observed at this point. Thus, we need to take the integral over half of the spherically distributed radial $\vec{k}$ directions, only:

$$\vec{E}_1(\vec{r}) = \frac{i\hat{E}_0}{2^3 \pi^2 k} \frac{\omega^2}{c_0^2} \iint_{\text{half of sphere} |\vec{k}_1|=k} \hat{\varepsilon} F(\vec{k}_1 - \vec{k}_0) \left[ (\vec{e}_0 \cdot \vec{e}_1^s) \vec{e}_1^s + (\vec{e}_0 \cdot \vec{e}_1^p) \vec{e}_1^p \right] \exp(i\vec{k}_1 \vec{r}) d^2 k_1. \tag{7}$$

Each plane wave can be viewed as a linear combination of two polarization components $\vec{e}_1^s$ and $\vec{e}_1^p$ corresponding to s- and p-polarization with respect to the plane given by the scattered vector $\vec{k}_1$ and by a z-direction $\vec{e}_z$ which in principle can be chosen arbitrarily. For particular structures considered here, like scattering films, it makes sense to choose the interface normal as the z-direction. Integration is performed over half of the sphere $|\vec{k}_1| = k$ since all allowed k-vectors need to be equal in length to the excitation wave vector. To understand the scattering itself one should look at the full sphere as in that case different observation points are of interest. The amplitudes of plane waves with different propagation directions $\vec{k}_1$ differ by the projection of the incident polarization vector $\vec{e}_0$ on the corresponding polarization vector $\vec{e}_1^s$ or $\vec{e}_1^p$, $\vec{e}_0 \cdot \vec{e}_1^s$ or $\vec{e}_0 \cdot \vec{e}_1^p$. This is not surprising as, e.g., the electric field of a current source oriented in a particular direction $\vec{e}_0$ will show the same dependence in the far field. Interestingly, the direction-dependent amplitudes depend moreover on the term $\hat{\varepsilon} F(\vec{k}_1 - \vec{k}_0)$; something we will visualize in the next section.

From the arbitrary z-direction introduced above and given by $\vec{e}_z$, half spaces can be defined, the limiting planes of which are perpendicular to the z-direction. For a finite scattering volume the power scattered into any such half space which does not contain the scatterer is

$$P_{1,z} = \iint \frac{1}{2} \text{Re}(\vec{E}_1 \times \vec{H}_1^*) \cdot \vec{e}_z dx dy = \frac{I_0}{2^4 \pi^2 k^2} \frac{\omega^4}{c_0^4} \iint_{\text{half of sphere} |\vec{k}_1|=k} |\hat{\varepsilon} F(\vec{k}_1 - \vec{k}_0)|^2 \left[ (\vec{e}_0 \cdot \vec{e}_1^s)^2 + (\vec{e}_0 \cdot \vec{e}_1^p)^2 \right] d^2 k_1,$$

(8)

where $I_0 = \bar{\varepsilon}^{1/2} |\hat{E}_0|^2 /(2Z_0)$ is the intensity of the incident wave and $Z_0 = (\mu_0/\varepsilon_0)^{1/2}$ is the vacuum wave impedance. Here, the $\vec{e}_z$-direction is chosen such that it points into the half space not containing the scatterer, the scattered power defined above is thus non-negative. All interference terms vanish as we integrate along the entire x-y-plane perpendicular to the z-direction and each plane wave contributes to the (spatial) spectral power in its propagation direction $\vec{k}_1$ with its intensity $\propto |\hat{\varepsilon} F(\vec{k}_1 - \vec{k}_0)|^2$. This vanishing of the interference is also the reason that power scattered in a certain direction is simply proportional to the squared absolute value of the corresponding field.

## 3. Visual representation of scattering (Ewald sphere)

We now want to discuss the scattering directions in more detail. "Direction" refers here to the propagation direction of the half space plane waves $\vec{k}_1$. A discussion of scattering directions in terms of solid angles which can be relevant for experimental situations can be found in Appendix B.

As mentioned, $\hat{\varepsilon} F(\vec{k})$ is the Fourier transform of the permittivity perturbation $\hat{\varepsilon} f(\vec{r})$. $\hat{\varepsilon} F(\vec{k} - \vec{k}_0)$ is the Fourier transform shifted by $\vec{k}_0$ in $k$-space. Finally, $\hat{\varepsilon} F(\vec{k}_1 - \vec{k}_0)$ is the shifted transform which is then evaluated on the sphere $|\vec{k}_1| = k$. Figure 1(a) illustrates the described evaluation as an intersection of the shifted Fourier-transformed permittivity perturbation with the sphere. Small absolute values of $\hat{\varepsilon} F(\vec{k}_1 - \vec{k}_0)$ correspond to small scattering amplitudes and, as a result, to weak scattering in that directions. Large values correspond to strong scattering. Shifting the coordinate system of the illustration in Figure 1(a) results in the equivalent Ewald sphere construction known from X-ray scattering [17] and depicted in Figure 1(b). This is a more useful representation as different incidence angles and frequencies can be considered by shifting and scaling the Ewald sphere, respectively, without changes in the Fourier transform of the perturbation. Instead of the perturbation's Fourier transform $\hat{\varepsilon} F(\vec{k}_1 - \vec{k}_0)$ (or more precisely its absolute value as complex values are hard to visualize) also its absolute value squared $|\hat{\varepsilon} F(\vec{k}_1 - \vec{k}_0)|^2$ can be used whenever scattering power is of more interest.

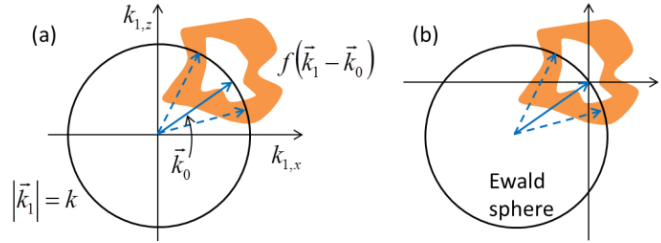

Figure 1. (a) Illustration of the shifted Fourier transform of the permittivity perturbation $\hat{\varepsilon} F(\vec{k}_1 - \vec{k}_0)$ and the sphere $|\vec{k}_1| = k$ at which $\hat{\varepsilon} F(\vec{k}_1 - \vec{k}_0)$ is evaluated. The dashed blue vectors show the directions $\vec{k}_1$ into which the scattering is strong because $\hat{\varepsilon} F(\vec{k}_1 - \vec{k}_0)$ has large (absolute) values there. Note that the original center of $\hat{\varepsilon} F(\vec{k}_1)$ is now lying on the sphere because $|\vec{k}_0| = k$. (b) Same case as shown in (a) however with the coordinate system shifted by $\vec{k}_0$. Here, $\hat{\varepsilon} F$ is not shifted but the sphere is, cutting the origin and having its center at $-\vec{k}_0$. This description is known from X-ray scattering where the sphere is called the Ewald sphere.

## 4. First-order scattering by infinitely extending perturbation / film

For a scattering structure infinitely extending into the *x*- and/or *y*-direction the scattered power is obviously infinite. As an example of such a structure we are now considering a film geometry with finite extension in the *z*-direction. Then, we can normalize the scattered power to obtain a power density defined as power per cross sectional area. For that, we take a representative volume limited to size $l$ in *x*- and *y*-direction. The representative volume should be larger than the correlation length in the considered disordered structure. In that case it can be assumed that there are no phase correlations with the rest of the structure, which is necessary for the sample to be representative of the whole structure. Let the permittivity perturbation in the sample volume be given by $\hat{\varepsilon} f_l(\vec{r})$. In the *z*-direction the perturbation is limited by the film thickness. The power scattered by this representative volume is given by (8) where only $\hat{\varepsilon} F(\vec{k})$ must be replaced by the corresponding Fourier transform $\hat{\varepsilon} F_l(\vec{k})$.

Dividing the result by the cross sectional area $l^2$ yields the power scattered per scattering area. If the statistical properties of the scattering structure do not change along the infinite extensions, the power scattered per scattering area is the same everywhere and equals the scattered power density.

The power density of the backscattered light can also be regarded as the Poynting vector averaged along the infinite extension dimensions and is given by:

$$\overline{S}_{1,z} = \frac{1}{2}\overline{\mathrm{Re}(\vec{E}_1 \times \vec{H}_1^*)} \cdot \vec{e}_z = \frac{I_0}{2^4 \pi^2 k^2} \frac{\omega^4}{c_0^4} \iint_{\text{half of sphere}|\vec{k}_1|=k} \frac{\left|\hat{\varepsilon} F_l(\vec{k}_1 - \vec{k}_0)\right|^2}{l^2} \left[(\vec{e}_0 \cdot \vec{e}_1^s)^2 + (\vec{e}_0 \cdot \vec{e}_1^p)^2\right] d^2 k_1. \tag{9}$$

Eq. (9) directly shows that the representative volume must be large enough to capture the statistical properties, otherwise scattering directions would depend on $l$. With Parseval's theorem it can be shown that $\overline{S}_{1,z}$ does not depend on $l$ as long as the spectral features do not change, which is a prerequisite, anyway.

In many real setups, the scattering film is not a simple perturbation inside a homogeneous material. Instead, we now assume that there is air above the film. For the material below the film we assume that it does not reflect any light. This is achieved by either filling the lower half space with a material of the same refractive index as the mean of the scattering film, or placing a non-reflecting absorber below the film. This asymmetric material arrangement mimics a typical setup e.g. for a structural color film.

While Eq. (9) gives us the full first-order scattering inside the film, the scattering into air outside the film is subject to additional angle-dependent Fresnel reflection at the surface. Also, the initial Fresnel reflection of the incident wave when entering the film has not been considered, yet. The interface effects for the case of normal incidence can be included in Eq. (9) as

$$\overline{S}_{1,z}^{air} = I_0^{air} R_{af} + \frac{I_0^{air} T_{af}}{(4\pi)^2 k^2 l^2} \frac{\omega^4}{c_0^4} \iint_{\text{half of sphere}|\vec{k}_1|=k} \left|\hat{\varepsilon} F_l(\vec{k}_1 - \vec{k}_0)\right|^2 \left[T_{fa}^s (\vec{e}_0 \cdot \vec{e}_1^s)^2 + T_{fa}^p (\vec{e}_0 \cdot \vec{e}_1^p)^2\right] d^2 k_1, \tag{10}$$

where $I_0^{air}$ is the intensity of the incident wave in air. $R_{af} = (\overline{\varepsilon}^{1/2} - 1)^2 / (\overline{\varepsilon}^{1/2} + 1)^2$ is the Fresnel reflectivity at the air-film interface for normal incidence and $T_{af} = 1 - R_{af}$ is the according transmissivity. The direction-dependent transmissivities through the same interface for light scattered from the lower into the upper half space are

$T^s_{fa} = 1 - \left|\left(k^{air}_{1,z} - k_{1,z}\right)/\left(k^{air}_{1,z} + k_{1,z}\right)\right|^2$ for s-polarization and $T^p_{fa} = 1 - \left|\left(\bar{\varepsilon}k^{air}_{1,z} - k_{1,z}\right)/\left(\bar{\varepsilon}k^{air}_{1,z} + k_{1,z}\right)\right|^2$ for p-polarization, where $k_{1,z} = \left(\left|\vec{k}_1\right|^2 - k^2_{1,\rho}\right)^{1/2}$ and $k^{air}_{1,z} = \left[(\omega/c_0)^2 - k^2_{1,\rho}\right]^{1/2}$ are the z-components of the wave vectors of the scattered light in the film and in air, respectively, where the tangential component is conserved at the transition between the media.

## 5. Discussion

The sum of the first two scattering orders, $\vec{E}_0 + \vec{E}_1$, is generally referred to as first-order Born approximation [16]. For weak perturbations the obtained result is identical with the Rayleigh-Gans approximation which splits the scattering volume in small Rayleigh scatterers and adds their scattering in the far field [18].

The advantage of the derivation presented here is a clear mathematical presentation of the Ewald sphere which allows direct calculation of the amplitudes or intensities of the scattered plane waves. The Ewald sphere also gives a very useful geometrical presentation of the excited scattering. Additionally, the Taylor-expansion approach leads to the possibility to consider excitations of higher orders as a recursive process, where second-order scattering is the scattering of the waves generated by the first-order scattering. The same perturbation function in reciprocal space is considered and just a new orientation of the Ewald sphere is applied. In situations with multiple scattering, the individual scattering events might be well describable by the first-order term. In this case, our approach or the Ewald sphere construction in general could be a part of a more complex calculation scheme where it describes the individual scattering events.

## 6. Conclusions

We have presented the detailed derivation of a scattering theory based on the first-order approximation and the Ewald sphere construction. Our formulation allows the quantitative calculation of the fields and power scattered at arbitrary geometries as well as the reflectivity of scattering films. Possible Fresnel reflections at the film interface are easily included. Since the approach focuses on the first-order scattering term it is, as a full scattering approach, limited to weakly scattering structures. We expect that our approach can find broad application in the field of disordered photonics and that it can help significantly cutting the computational effort in modeling the scattering of disordered photonic structures.

**Acknowledgements.** This work was supported by the German Research Foundation (DFG) via priority program SPP 1839 "Tailored Disorder" and via the Collaborative Research Centre SFB 986 "Tailor-Made Multi-Scale Materials Systems (M³)".

## Appendix A: Derivation of the Green's function

The Green's function for the electric field $\mathbf{G}$ is derived from

$$\nabla \times \nabla \times \mathbf{G}(\vec{r}) - k^2 \mathbf{G}(\vec{r}) = \mathbf{I}\delta(\vec{r}) \quad \Leftrightarrow \quad \nabla(\nabla \cdot \mathbf{G}(\vec{r})) - \nabla^2 \mathbf{G}(\vec{r}) - k^2 \mathbf{G}(\vec{r}) = \mathbf{I}\delta(\vec{r}), \tag{A1}$$

where $\delta$ is the Dirac delta function and $\mathbf{I}$ the unit matrix. The differential operators are applied column vise. A particular solution can be determined in Fourier space [19]:

$$i\vec{k}\left(i\vec{k}\cdot\mathbf{G}_{inh}(\vec{k})\right)-\left(i\vec{k}\cdot i\vec{k}\right)\mathbf{G}_{inh}(\vec{k})-k^2\mathbf{G}_{inh}(\vec{k})=\mathbf{I}$$
$$\Leftrightarrow \quad -(\vec{k}\otimes\vec{k})\mathbf{G}_{inh}(\vec{k})+|\vec{k}|^2\mathbf{G}_{inh}(\vec{k})-k^2\mathbf{G}_{inh}(\vec{k})=\mathbf{I} \tag{A2}$$
$$\Rightarrow \quad \mathbf{G}_{inh}(\vec{k})=-\left[\vec{k}\otimes\vec{k}+\left(k^2-|\vec{k}|^2\right)\mathbf{I}\right]^{-1}=-\frac{\mathrm{adj}\left[\vec{k}\otimes\vec{k}+\left(k^2-|\vec{k}|^2\right)\mathbf{I}\right]}{\det\left[\vec{k}\otimes\vec{k}+\left(k^2-|\vec{k}|^2\right)\mathbf{I}\right]}.$$

It can be shown that

$$\mathrm{adj}\left[\vec{k}\otimes\vec{k}+\left(k^2-|\vec{k}|^2\right)\mathbf{I}\right]=\left(k^2-|\vec{k}|^2\right)\left[k^2\mathbf{I}-\vec{k}\otimes\vec{k}\right]$$
$$\det\left[\vec{k}\otimes\vec{k}+\left(k^2-|\vec{k}|^2\right)\mathbf{I}\right]=k^2\left(k^2-|\vec{k}|^2\right)^2 \tag{A3}$$

so that Green's function becomes

$$\mathbf{G}_{inh}(\vec{k})=\frac{k^2\mathbf{I}-\vec{k}\otimes\vec{k}}{k^2\left(|\vec{k}|^2-k^2\right)}=\frac{1}{k^2\left(|\vec{k}|^2-k^2\right)}\begin{pmatrix} k^2-k_x^2 & -k_xk_y & -k_xk_z \\ -k_xk_y & k^2-k_y^2 & -k_yk_z \\ -k_xk_z & -k_yk_z & k^2-k_z^2 \end{pmatrix}. \tag{A4}$$

We now define the $z$-component of a $k$-vector of length $k$ as $\tilde{k}_z:=\left(k^2-|\vec{k}_\rho|^2\right)^{1/2}$ (note that $|\vec{k}|^2-k^2=(k_z-\tilde{k}_z)(k_z+\tilde{k}_z)$) and the unit polarization vectors (with respect to $z$-axis) [19]

$$\vec{e}_\pm^s:=\frac{\begin{pmatrix}\vec{k}_\rho \\ \pm\tilde{k}_z\end{pmatrix}\times\vec{e}_z}{\left|\begin{pmatrix}\vec{k}_\rho \\ \pm\tilde{k}_z\end{pmatrix}\times\vec{e}_z\right|}=\frac{1}{|\vec{k}_\rho|}\begin{pmatrix}k_y \\ -k_x \\ 0\end{pmatrix},\quad \vec{e}_\pm^p:=\frac{\begin{pmatrix}\vec{k}_\rho \\ \pm\tilde{k}_z\end{pmatrix}\times\vec{e}_\pm^s}{\left|\begin{pmatrix}\vec{k}_\rho \\ \pm\tilde{k}_z\end{pmatrix}\times\vec{e}_\pm^s\right|}=\frac{1}{k|\vec{k}_\rho|}\begin{pmatrix}\pm k_x\tilde{k}_z \\ \pm k_y\tilde{k}_z \\ -|\vec{k}_\rho|^2\end{pmatrix} \tag{A5}$$

with the tensor product properties

$$\frac{\vec{e}_+^s\otimes\vec{e}_+^s}{2\tilde{k}_z(k_z-\tilde{k}_z)}-\frac{\vec{e}_-^s\otimes\vec{e}_-^s}{2\tilde{k}_z(k_z+\tilde{k}_z)}=\frac{1}{\left(|\vec{k}|^2-k^2\right)|\vec{k}_\rho|^2}\begin{pmatrix} k_y^2 & -k_xk_y & 0 \\ -k_xk_y & k_x^2 & 0 \\ 0 & 0 & 0 \end{pmatrix}$$

$$\frac{\vec{e}_+^p\otimes\vec{e}_+^p}{2\tilde{k}_z(k_z-\tilde{k}_z)}-\frac{\vec{e}_-^p\otimes\vec{e}_-^p}{2\tilde{k}_z(k_z+\tilde{k}_z)}=\frac{1}{k^2\left(|\vec{k}|^2-k^2\right)|\vec{k}_\rho|^2}\begin{pmatrix} k_x^2\tilde{k}_z^2 & k_xk_y\tilde{k}_z^2 & -k_xk_z|\vec{k}_\rho|^2 \\ k_xk_y\tilde{k}_z^2 & k_y^2\tilde{k}_z^2 & -k_yk_z|\vec{k}_\rho|^2 \\ -k_xk_z|\vec{k}_\rho|^2 & -k_yk_z|\vec{k}_\rho|^2 & |\vec{k}_\rho|^4 \end{pmatrix}, \tag{A6}$$

making it possible to express Green's function in the form

$$\mathbf{G}_{inh}(\vec{k})=\frac{k^2\mathbf{I}-\vec{k}\otimes\vec{k}}{k^2(k_z-\tilde{k}_z)(k_z+\tilde{k}_z)}=\frac{\vec{e}_+^s\otimes\vec{e}_+^s+\vec{e}_+^p\otimes\vec{e}_+^p}{2\tilde{k}_z(k_z-\tilde{k}_z)}-\frac{\vec{e}_-^s\otimes\vec{e}_-^s+\vec{e}_-^p\otimes\vec{e}_-^p}{2\tilde{k}_z(k_z+\tilde{k}_z)}+\Gamma(\vec{k}) \tag{A7}$$

$\boldsymbol{\Gamma}(\vec{k})$ is the constant matrix

$$\boldsymbol{\Gamma}(\vec{k}) = \frac{-1}{k^2} \begin{pmatrix} 0 & 0 & 0 \\ 0 & 0 & 0 \\ 0 & 0 & 1 \end{pmatrix} \tag{A8}$$

and thus, represents a Dirac delta function in real space at the origin of the source. In the following we neglect it as it is irrelevant for emitted power calculations.

The inverse Fourier transform with respect to $z$ is given by

$$\mathbf{G}_{inh}(\vec{k}_\rho, z) = \frac{i}{4\tilde{k}_z}\left[\vec{e}_+^s \otimes \vec{e}_+^s + \vec{e}_+^p \otimes \vec{e}_+^p\right]\exp(i\tilde{k}_z z)\operatorname{sgn}(z) - \frac{i}{4\tilde{k}_z}\left[\vec{e}_-^s \otimes \vec{e}_-^s + \vec{e}_-^p \otimes \vec{e}_-^p\right]\exp(-i\tilde{k}_z z)\operatorname{sgn}(z) \tag{A9}$$

where $\operatorname{sgn}(z)$ is the sign function. This is a superposition of plane waves (with a sign change at $z=0$) as can be seen from the inverse Fourier transform

$$\mathbf{G}_{inh}(\vec{r}) = \frac{1}{(2\pi)^2}\iint \mathbf{G}_{inh}(\vec{k}_\rho, z)\exp(i\vec{k}_\rho \vec{\rho})d^2k_\rho. \tag{A10}$$

This is a fully vectorial equivalence of the Weyl's angular spectrum representation [20] of a point source. The obtained Green's function is not a physical solution as it also contains waves traveling towards the source (incoming waves) which do not exist for a physical point source. But the general solution of the linear, inhomogeneous differential equation (A1) is the superposition of a particular solution which is given by (A9) and a weighted solution of the corresponding homogeneous differential equation

$$\nabla \times \nabla \times \mathbf{G}_{hom}(\vec{r}) - k^2 \mathbf{G}_{hom}(\vec{r}) = 0 \quad \Leftrightarrow \quad \nabla(\nabla \cdot \mathbf{G}_{hom}(\vec{r})) - \nabla^2 \mathbf{G}_{hom}(\vec{r}) - k^2 \mathbf{G}_{hom}(\vec{r}) = 0$$
$$\left[\vec{k} \otimes \vec{k} + \left(k^2 - |\vec{k}|^2\right)\mathbf{I}\right]\mathbf{G}_{hom}(\vec{k}) = 0. \tag{A11}$$

The weight is chosen such that the solution satisfies all boundary or/and initial conditions. Here, the specific solution (A9) consists of waves traveling away from the source located at the origin, $\propto \exp(i\tilde{k}_z z)$ for $z > 0$ and $\propto \exp(-i\tilde{k}_z z)$ for $z < 0$, and of waves traveling towards the source, $\propto \exp(-i\tilde{k}_z z)$ for $z > 0$ and $\propto \exp(i\tilde{k}_z z)$ for $z < 0$. The waves traveling towards the source are unphysical and the weight should be chosen such that they vanish. With

$$\mathbf{G}_{hom}(\vec{k}_\rho, z) = \frac{i}{4\tilde{k}_z}\left[\vec{e}_+^s \otimes \vec{e}_+^s + \vec{e}_+^p \otimes \vec{e}_+^p\right]\exp(i\tilde{k}_z z) + \frac{i}{4\tilde{k}_z}\left[\vec{e}_-^s \otimes \vec{e}_-^s + \vec{e}_-^p \otimes \vec{e}_-^p\right]\exp(-i\tilde{k}_z z)$$

$$\mathbf{G}_{hom}(\vec{k}) = \frac{i\pi}{2\tilde{k}_z}\left[\vec{e}_+^s \otimes \vec{e}_+^s + \vec{e}_+^p \otimes \vec{e}_+^p\right]\delta(k_z - \tilde{k}_z) + \frac{i\pi}{2\tilde{k}_z}\left[\vec{e}_-^s \otimes \vec{e}_-^s + \vec{e}_-^p \otimes \vec{e}_-^p\right]\delta(k_z + \tilde{k}_z) \tag{A12}$$

we obviously obtain the physical Green's function

$$\mathbf{G} = \mathbf{G}_{inh} + \mathbf{G}_{hom}$$

$$\mathbf{G}(\vec{k}_\rho, z) = \frac{i}{2\tilde{k}_z}\left[\vec{e}_+^s \otimes \vec{e}_+^s + \vec{e}_+^p \otimes \vec{e}_+^p\right]\exp(i\tilde{k}_z z)\sigma(z) + \frac{i}{2\tilde{k}_z}\left[\vec{e}_-^s \otimes \vec{e}_-^s + \vec{e}_-^p \otimes \vec{e}_-^p\right]\exp(-i\tilde{k}_z z)\sigma(-z)$$

$$\mathbf{G}(\vec{k}) = \left(\frac{i\pi}{2\tilde{k}_z}\delta(k_z - \tilde{k}_z) + \frac{1}{2\tilde{k}_z(k_z - \tilde{k}_z)}\right)\left[\vec{e}_+^s \otimes \vec{e}_+^s + \vec{e}_+^p \otimes \vec{e}_+^p\right]$$
$$+ \left(\frac{i\pi}{2\tilde{k}_z}\delta(k_z + \tilde{k}_z) - \frac{1}{2\tilde{k}_z(k_z + \tilde{k}_z)}\right)\left[\vec{e}_-^s \otimes \vec{e}_-^s + \vec{e}_-^p \otimes \vec{e}_-^p\right]$$

(A13)

with the Heaviside step function $\sigma(z)$ and no incoming waves. It contains only outgoing plane waves defined in the half space starting or ending from the point source, respectively. The amplitude of the outgoing half space plane waves is double of the added plane waves. Basically, the added plane waves cancel the unphysical incoming waves and doubled the outgoing waves. This is similar to results obtained for the scalar wave equation [21]. The Green's function can now be written as:

$$\mathbf{G}(\vec{k}) = \frac{i\pi}{2k}\delta(|\vec{k}| - k)\left[\vec{e}^s \otimes \vec{e}^s + \vec{e}^p \otimes \vec{e}^p\right] + \frac{1}{\left(|\vec{k}|^2 - k^2\right)}\left[\vec{e}^s \otimes \vec{e}^s + \frac{k_z + \tilde{k}_z}{2\tilde{k}_z}\vec{e}_+^p \otimes \vec{e}_+^p - \frac{k_z - \tilde{k}_z}{2\tilde{k}_z}\vec{e}_-^p \otimes \vec{e}_-^p\right],$$

(A14)

introducing

$$\vec{e}^s := \frac{1}{|\vec{k}_\rho|}\begin{pmatrix} k_y \\ -k_x \\ 0 \end{pmatrix} \quad \text{and} \quad \vec{e}^p := \frac{1}{k|\vec{k}_\rho|}\begin{pmatrix} k_x k_z \\ k_y k_z \\ -|\vec{k}_\rho|^2 \end{pmatrix}.$$

(A15)

We insert Green's function (A14) into Eq. (5) and perform the integration along the radial $\vec{k}$-direction to obtain Eq. (7). As discussed before, Green's function is expressed as a superposition of plane waves emitted away from the source. The Dirac delta in Eq. (A14) enforces that only wave vectors with $|\vec{k}| = k$ represent the plane waves, the other term in (A14) ensures that the waves exist only in a half space. After integration, the wave vectors are limited to $|\vec{k}| = k$ which is the characteristic of plane waves inside a medium with permittivity $\bar{\varepsilon}$. Since the plane waves exist in a half space only, at any observation point one observes only those waves emitted towards the observation point, which comprise half of the sphere $|\vec{k}| = k$. Following the discussion above, the amplitudes of the waves are two times the amplitudes of the Dirac delta in Eq. (A14).

## Appendix B: Scattering in a Certain Solid Angle

In this part we want to discuss the scattering directions. For that we start with the first-order scattered field outside the perturbation derived previously and given by (7):

$$\vec{E}_1(\vec{r}) = \frac{i\hat{E}_0}{2^3\pi^2 k}\frac{\omega^2}{c_0^2}\iint\limits_{\text{half of sphere}|\vec{k}_1|=k}\hat{\varepsilon}\,F(\vec{k}_1 - \vec{k}_0)\left[(\vec{e}_0 \cdot \vec{e}_1^s)\vec{e}_1^s + (\vec{e}_0 \cdot \vec{e}_1^p)\vec{e}_1^p\right]\exp(i\vec{k}_1\vec{r})d^2k_1. \quad \text{(A16)}$$

The electric field $\vec{E}_1(\vec{r})$ is represented as a superposition of half space plane waves emitted by the permittivity perturbation. Each plane wave has a certain propagation direction $\vec{k}_1$. However, these plane waves exist in the entire half space. So, the power scattered into a small solid angle is given by contributions of plane waves with different directions and even by interference contributions of different plane waves.

We will now show that far away from the scatterer things significantly simplify. Assume the scatterer has a finite size $V'$, meaning $f(\vec{r}') = 0$ for $\vec{r}'$ outside this volume, and its center is located at the origin of our coordinate system. By explicitly writing down the Fourier integral the scattered field becomes

$$\vec{E}_1(\vec{r}) = \frac{i\hat{E}_0}{8\pi^2 k} \frac{\omega^2}{c_0^2} \iint\limits_{\text{half of sphere } |\vec{k}_1|=k} \iiint\limits_{V'} \hat{\varepsilon} f(\vec{r}') \exp\left(-i(\vec{k}_1 - \vec{k}_0)\vec{r}'\right) d^3 r' \left[ (\vec{e}_0 \cdot \vec{e}_1^s)\vec{e}_1^s + (\vec{e}_0 \cdot \vec{e}_1^p)\vec{e}_1^p \right] \exp(i\vec{k}_1 \vec{r}) d^2 k_1$$

$$= \frac{i\hat{E}_0}{8\pi^2 k} \frac{\omega^2}{c_0^2} \iiint\limits_{V'} \hat{\varepsilon} f(\vec{r}') \exp(-i\vec{k}_0 \vec{r}') \iint\limits_{\text{half of sphere } |\vec{k}_1|=k} \left[ (\vec{e}_0 \cdot \vec{e}_1^s)\vec{e}_1^s + (\vec{e}_0 \cdot \vec{e}_1^p)\vec{e}_1^p \right] \exp(i\vec{k}_1(\vec{r} - \vec{r}')) d^2 k_1 d^3 r'.$$

(A17)

Furthermore, analogously to the scalar Green's function considered by Weyl [20] for the case of a Hertzian dipole we know that

$$\iint\limits_{\text{half of sphere } |\vec{k}_1|=k} \left[ (\vec{e}_0 \cdot \vec{e}_1^s)\vec{e}_1^s + (\vec{e}_0 \cdot \vec{e}_1^p)\vec{e}_1^p \right] \frac{\exp(i\vec{k}_1(\vec{r} - \vec{r}'))}{k} d^2 k_1 = 2\pi\sqrt{\varepsilon} \frac{\exp(ik|\vec{r} - \vec{r}'|)}{|\vec{r} - \vec{r}'|} (\vec{e}_0 \cdot \vec{e}_\alpha) \vec{e}_\alpha$$

(A18)

with the unit polar polarization vector

$$\vec{e}_\alpha = \frac{|\vec{r} - \vec{r}'|}{|(\vec{r} - \vec{r}') - [\vec{e}_0 \cdot (\vec{r} - \vec{r}')]\vec{e}_0|} \left\{ \vec{e}_0 - \left[ \vec{e}_0 \cdot \frac{\vec{r} - \vec{r}'}{|\vec{r} - \vec{r}'|} \right] \frac{\vec{r} - \vec{r}'}{|\vec{r} - \vec{r}'|} \right\}.$$

(A19)

This result can be interpreted analogously to a Hertzian dipole oriented in the $\vec{e}_0$-direction emitting spherical waves in the far field, the amplitudes of which are weighted by the sine $\sin(\alpha) = \vec{e}_0 \cdot \vec{e}_\alpha$ of the polar angle $\alpha$.

Far away from the scatterer we can approximate $\vec{r} - \vec{r}'$ by $\vec{r} - \vec{r}' = \vec{r}$. For the more sensitive phase term $\exp(ik|\vec{r} - \vec{r}'|)$ we use the more precise approximation $|\vec{r} - \vec{r}'| \approx |\vec{r}| - \vec{e}_r \cdot \vec{r}'$ where $\vec{e}_r = \vec{r}/|\vec{r}|$ is the unit vector in the $\vec{r}$-direction [16]. We obtain

$$\vec{E}_1(\vec{r}) = \frac{i\sqrt{\varepsilon}\hat{E}_0}{4\pi} \frac{\omega^2}{c_0^2} \frac{\exp(ik|\vec{r}|)}{|\vec{r}|} (\vec{e}_0 \cdot \vec{e}_\alpha) \vec{e}_\alpha \iiint\limits_{V'} \hat{\varepsilon} f(\vec{r}') \exp\left(i(\vec{k}_0 - k\vec{e}_r)\vec{r}'\right) d^3 r'$$

$$= \frac{i\sqrt{\varepsilon}\hat{E}_0}{4\pi} \frac{\omega^2}{c_0^2} \hat{\varepsilon} F(k\vec{e}_r - \vec{k}_0) \frac{\exp(ik|\vec{r}|)}{|\vec{r}|} (\vec{e}_0 \cdot \vec{e}_\alpha) \vec{e}_\alpha.$$

(A20)

Far away the scattered field is like the field of a point source but weighted by $\hat{\varepsilon} F(k\vec{e}_r - \vec{k}_0)$. Considering that the amplitudes of the plane waves are proportional to $\hat{\varepsilon} F(\vec{k}_1 - \vec{k}_0)$ we see that at larger distances the plane wave with the propagation direction $k\vec{e}_r$ – along the connection scatterer center – observation point – dominates. The power scattered into $\vec{e}_r$- direction is proportional to $\left|\hat{\varepsilon}\, F(k\vec{e}_r - \vec{k}_0)\right|^2$. Thus, the power scattered into a certain cone of directions $\Omega$ can be evaluated in the reciprocal space as

$$P_{1,\Omega} = I_0 \frac{\omega^4}{c_0^4} \frac{1}{(4\pi)^2} \iint_{\Omega \text{ on } |\vec{k}_1|=k} \left|\hat{\varepsilon}\, F(k\vec{e}_k - \vec{k}_0)\right|^2 (\vec{e}_0 \cdot \vec{e}_\alpha)^2 \frac{d^2 k_1}{k^2},$$

(A21)

where the integration is performed over the part of the surface area of the Ewald sphere corresponding to the directions given by $\Omega$. Here, $\vec{e}_k = \vec{k}_1 / k$ is the equivalent to $\vec{e}_r$ in reciprocal space. For unpolarized light the term $(\vec{e}_0 \cdot \vec{e}_\alpha)^2$ can be replaced by $g(\theta) = (1 + \cos^2 \theta)/2$ where $\theta$ is the angle between the incident and the scattered $k$-vector. Eq. (A21) allows a quantitative prediction of angle distributions of scattered power.